\documentclass{article}

\PassOptionsToPackage{numbers, compress}{natbib}

\usepackage[preprint]{neurips_2024}

\usepackage[utf8]{inputenc} 
\usepackage[T1]{fontenc}    
\usepackage{hyperref}       
\usepackage{url}            
\usepackage{booktabs}       
\usepackage{amsfonts}       
\usepackage{nicefrac}       
\usepackage{microtype}      
\usepackage{xcolor}         
\usepackage{float}
 \usepackage{graphicx}
\usepackage{graphicx}   
\usepackage{subcaption} 
\usepackage[skip=0.5cm]{caption} 
\usepackage{svg}
\usepackage{amsmath}
\title{Starkindler: An Uncertainty Aware Objective for Photometric Redshift Estimation}

\vspace{-10pt} 
\author{%
  Raahul Singh \\
  Phaidra \\
  \texttt{rasalghul@phaidra.ai} \\
   \And
   Ashutosh Pandey \\
   Advanced Micro Devices (AMD) \\
   \texttt{ashutosh.pandey@amd.com} \\
}

\begin{document}

\maketitle

\vspace{-10pt} 
\begin{abstract}
\vspace{-10pt} 
Photometric Redshift is critical for analyzing astronomical objects, but existing ML methods often overlook the aleatoric uncertainties inherent in observed data. We introduce \textbf{Starkindler}, a novel training objective that explicitly incorporates observational errors into the model’s objective function, thereby directly accounting for aleatoric uncertainty. Unlike traditional probabilistic models that focus solely on epistemic uncertainty, Starkindler provides uncertainty estimates that are regularised by aleatoric uncertainty, and is designed to be more interpretable. We train a simple convolutional neural network (CNN) using data from Sloan Digital Sky Survey (SDSS)\cite{york2000sloan} and compare against the Photometric redshift estimates provided by SDSS. We show improvements in accuracy, calibration and reduction in predicted outlier rate. We also conduct an ablation study which confirms that excluding observational errors significantly degrades model performance, underscoring the importance of accounting for aleatoric uncertainty. Our results suggest that Starkindler not only enhances predictive performance but also provides interpretable uncertainty estimates, making it a robust tool for astronomical data analysis.
\end{abstract}

\section{Introduction}
\vspace{-10pt} 
Redshift is the increase in the wavelength of electromagnetic radiation emitted by distant astronomical objects due to their relative motion to Earth as well as the expansion of the universe. Redshift estimation is essential for measuring galactic distances, studying their evolution, constraining cosmological models, and enabling large-scale surveys, thereby providing critical insights into the structure and dynamics of the universe\cite{lahav2004}\cite{newman2022}. Redshift estimation via spectroscopy ($z_{spec}$) is considered the gold standard due to its high accuracy and reliability in measuring the redshift of astronomical objects\cite{masters2017complete}\cite{hasinger2018deimos}\cite{kriek2015mosfire}\cite{vanderwel2021lega}. However, spectroscopy is resource-intensive, requiring significant time and effort to obtain redshift measurements for individual galaxies. This makes it impractical for large surveys that aim to study vast populations of galaxies. An alternative method, Photometric redshift estimation ($z_{photo}$) utilizes multi-band photometric data to estimate redshifts. While $z_{photo}$ is more cost-effective and efficient, it is subject to higher uncertainties due to its reliance on approximations of the spectral energy distributions (SEDs) of galaxies\cite{newman2022}. The next generation of wide-field imaging surveys including the Rubin\cite{schmidt2020evaluation}, Euclid\cite{schirmer2022euclid} and Roman Space Telescope\cite{eifler2021cosmology} aim to observe billions of galaxies which necessitates improvements in $z_{photo}$ estimation techniques.

%
\section{Related Works}
\vspace{-5pt} 
    Several methods use $z_{spec}$ values as the baseline to develop models that can predict redshifts for galaxies based on their photometric data. These techniques can be broadly categorized into two main approaches: \textbf{Template Fitting} methods and \textbf{Machine Learning} methods. Template fitting methods establish the relationship between photometry and redshift by matching SEDs to a set of template SEDs. Some of the most widely used template fitting solutions are\cite{arnouts2011}\cite{Brammer_2008}\cite{Benítez_2000}. Machine learning techniques aim to directly map the relationship between photometry and redshift using supervised learning on a spectroscopic training set. Popular approaches include\cite{firth}\cite{10.1093/mnras/stac480}\cite{schuldt2021photometric}\cite{sun2023zephyr}. A key limitation of current template fitting and machine learning approaches is that they do not adequately account for the inherent uncertainties $zErr_{spec}$ in redshift estimation. Template-based methods use an informative prior, the underlying model of the full galaxy population to model $z_{photo}$. This requirement on the prior can be problematic when the template set is incomplete\cite{malz2022obtain}. On the other hand, current Machine Learning based estimates of uncertainty are difficult to interpret\cite{newman2022}, and since they do not include aleatoric uncertainty, are restricted to estimating epistemic uncertainty. This prevents these methods from being widely adopted by the scientific community.

\section{Method}
\vspace{-5pt} 
\subsection{Methodology}
\vspace{-5pt} 
Starkindler interprets the spectroscopic Redshift $z_{spec}$, and the redshift error ($zErr_{spec}$) (both provided by SDSS) as the mean ($\mu_{spec}$) and standard deviation ($\sigma_{spec}$) of the ground truth target Gaussian distribution for redshift at each data point in the dataset. This modeling choice aligns with common practices in astronomy, where errors in Redshift estimates are typically treated as normally distributed \cite{mandelbaum2018lsst}\cite{disanto2018photometric}\cite{oyaizu2008photometric}\cite{fernandezsoto2002error}, and thus includes aleatoric uncertainty in the training objective:
\begin{equation}
z_{spec} \sim \mathcal{N}(\mu_{spec}, \sigma_{spec}^2)
\end{equation}

Similarly, the model takes the 5 channel photometric image $X$ as input, and predict the mean ($\mu_{pred}$) and standard deviation ($\mu_{pred}$), parameterising a Gaussian distribution prediction for each image:
\begin{equation}
z_{pred} \sim \mathcal{N}(\mu_{pred}, \sigma_{pred}^2)
\end{equation}
We compute the Kullback-Leibler (KL) divergence between the true and predicted Gaussian distributions, which can be computed in closed form for two Gaussians: 
\begin{equation}
KL(z_{spec} \ || \ z_{pred}) = \log \frac{\sigma_{pred}}{\sigma_{spec}} + \frac{\sigma_{spec}^2 + (\mu_{spec} - \mu_{pred})^2}{2 \sigma_{pred}^2} - \frac{1}{2}
\end{equation}

The training objective is then the minimisation of this divergence.
\textbf{To our knowledge, Starkindler is the first method to incorporate $zErr$ directly into model training, thus accounting for aleatoric uncertainty in the observed data.} To show the efficacy of the model, we train a simple CNN which has two convolutional layers with 32 channels with a kernel size of 3 each. This serves as a feature extractor which feeds two MLP heads with three hidden layer of width 1024, 512 and 32 each. We ensure non-negative predictions by clamping the mean prediction at the lower end, and using a softplus activation for the standard deviation prediction.

\section{Experiments}
\vspace{-10pt} 
\begin{figure} [H]
  \centering
   \includegraphics[width=0.7\textwidth]{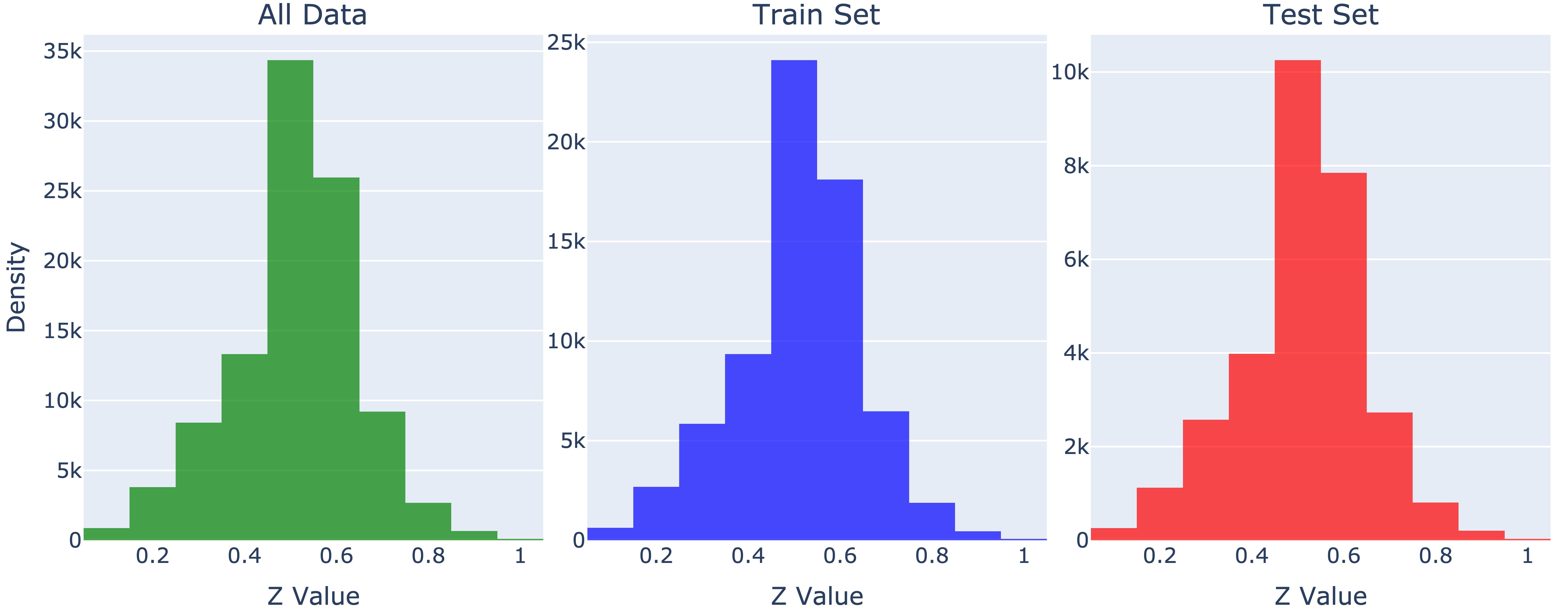}
    \vspace{-10pt} 
  \caption{Test Train Split.}
  \label{fig:1}
    \vspace{-10pt} 
\end{figure}

\subsection{Dataset}
\vspace{-5pt} 
We query the SDSS SkyServer archive\cite{neilsen2007sdss} (DR17\cite{abdurrouf2022seventeenth}) for the top 100,000 galaxies (ordered by object id) with $z_{spec}$ values between 0.1 and 1, and get the $z_{spec}$, $zErr_{spec}$, $z_{photo}$, $zErr_{photo}$, and galactic coordinates, and download each corresponding frame. Each galaxy image extracted as a 40x40 grid centered on its coordinates (RA, Dec), with data spanning the five $UGRIZ$ bands\cite{sichevskij2018applicability}. After filtering out data points with negative $zErr_{spec}$ or $zErr_{photo}$ values (which represent incorrect data as per SDSS), we retain 99,388 images. We have provided the query to get the exact dataset from SDSS, along with code to download and process the images from SDSS servers. We perform a stratified sampling based on binned z values at 0.1 intervals to create a 70/30 train-test split giving a train and test sets with 69,571 and 29,817 images respectively, while ensuring balanced representation across the redshift range and minimizing sampling bias (Figure \ref{fig:1}). The experiments were conducted on an M3 Pro Macbook Pro GPU, with 16GB of Unified Memory for 10 epochs.

\subsection{Baseline}
\vspace{-5pt} 
We compare Starkindler and the ablation model against the official SDSS photometric redshift estimates, which represent the most accurate baseline photometric redshift estimates available from SDSS\cite{beck2016photometric}. As an ablation study, we also train the exact same architecture using an Negative Log Likelihood (NLL) objective wherein the only the likelihood of the observed $z_{spec}$ is maximised, without considering the $zErr_{spec}$. 

\subsection{Evaluation Metrics}
\vspace{-5pt} 
\textbf{Redshift Prediction Accuracy}: We use the Mean Squared Error (MSE), Mean Absolute Error (MAE) of mean predictions, along with  scatter plots of $z_{pred}$ vs $z_{spec}$ as density contours with highlighted outliers, defined as instances where $|((\mu_{spec} - \mu_{pred}) / (1 + \mu_{spec}))| > 0.3$,  as used in \cite{li2023photometric}.

\textbf{Uncertainty Coverage and Calibration}: Evaluated through Expected Calibration Error (ECE), Probability Integral Transform (PIT) histograms and Coverage Probability Plots.

\section{Results}
\vspace{-5pt} 
The results demonstrate the effectiveness of Starkindler in enhancing both accuracy and uncertainty quantification:

\begin{table}[h]
    \centering
    \begin{tabular}{lccccc}
        \hline
        & \textbf{MSE} & \textbf{MAE} & \textbf{ECE} \\
        \textbf{Method} & \textbf{mean ± std} & \textbf{mean ± std} & \textbf{} \\
        \hline
        baseline & 0.0054 ± 0.0215 & 0.0449 ± 0.0582 & 7.46 \\
        starkindler & \textbf{0.0042 ± 0.0130} & \textbf{0.0441 ± 0.0479} & 6.10 \\
        ablation\_nll & 0.0127 ± 0.0319 & 0.0807 ± 0.0787 & \textbf{2.55} \\
        \hline
    \end{tabular}
    \caption{Performance metrics.}
    \label{tab:performance}
\end{table}
\begin{figure}[ht]
    \centering
    \vspace{-0.5cm} 
    \begin{subfigure}[b]{0.9\textwidth}
        \centering
        \includegraphics[width=\textwidth,height=0.2\textheight]{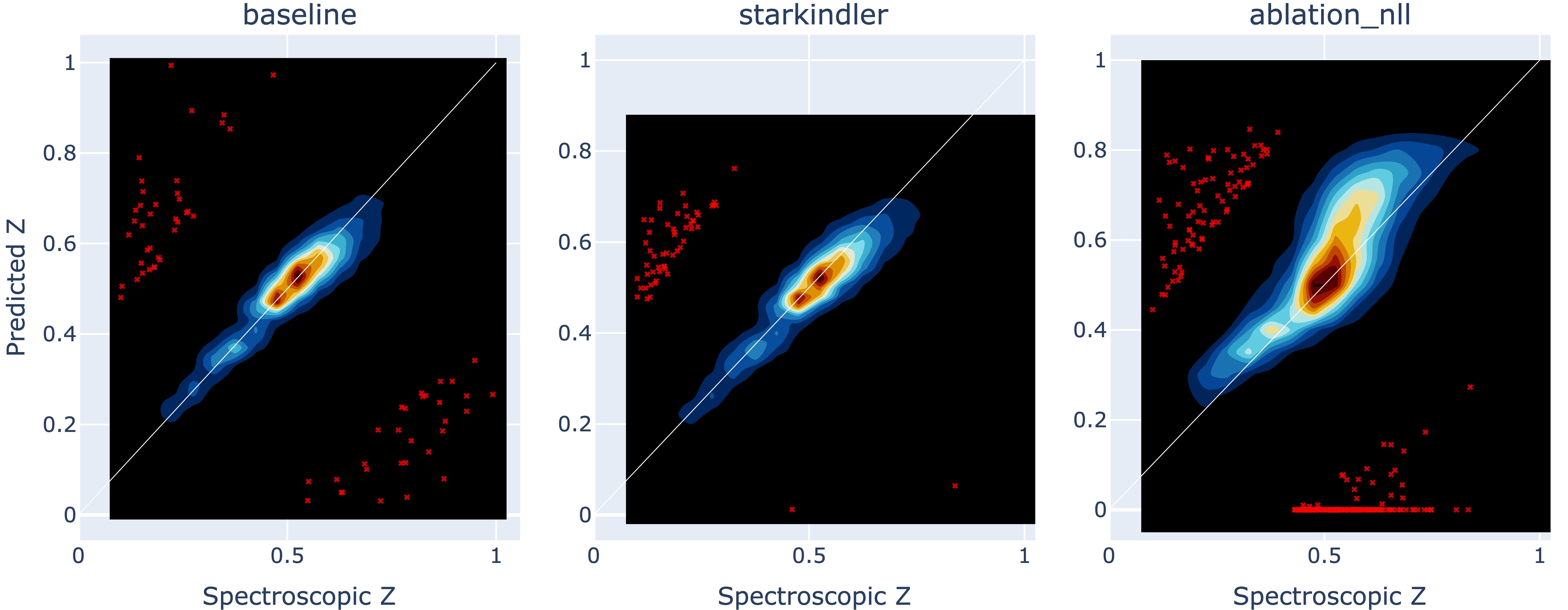} 
            \vspace{-10pt} 
        \caption{Predicted Z vs Spec Z Density plot with Outliers.}
        \label{fig:fig2a}
    \end{subfigure}

    \vspace{0.1cm} 
    \begin{subfigure}[b]{0.9\textwidth}
        \centering
        \includegraphics[width=\textwidth,height=0.2\textheight]{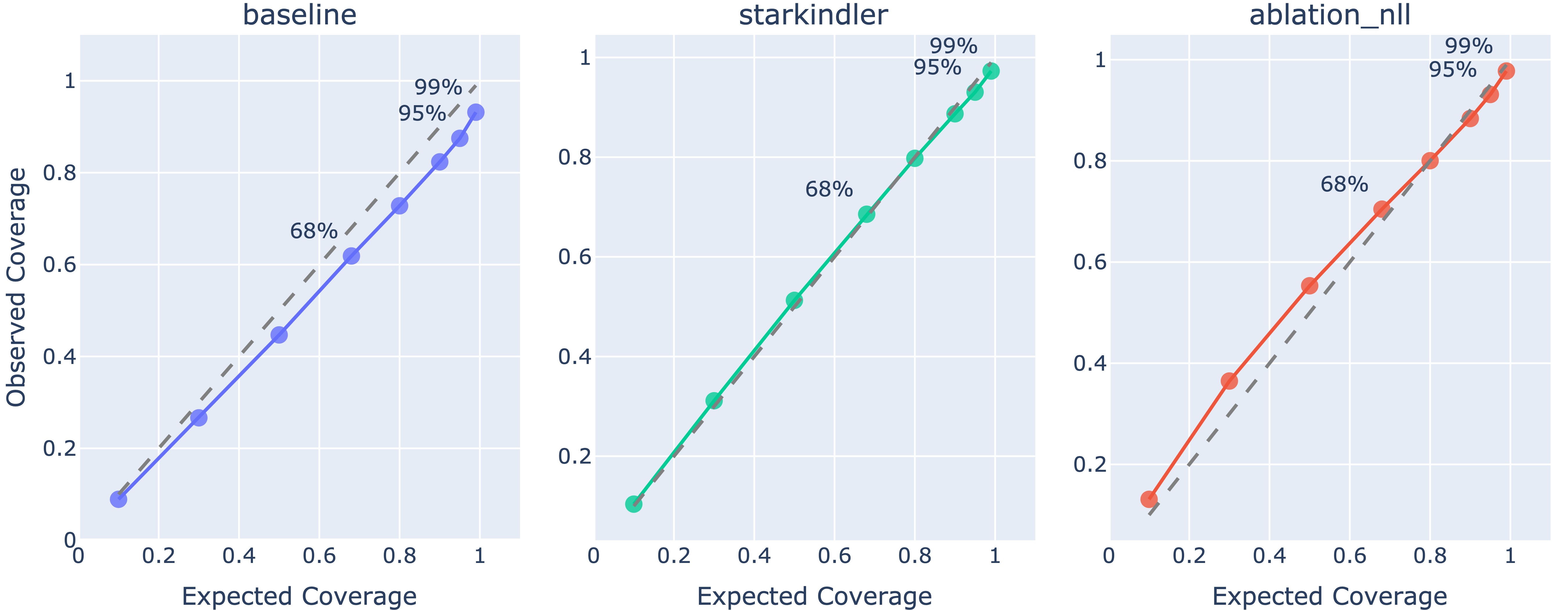} 
        \caption{Probability Coverage plots.}
        \label{fig:fig2b}
    \end{subfigure}
    
    \vspace{0.1cm} 
    \begin{subfigure}[b]{0.9\textwidth}
        \centering
        \includegraphics[width=\textwidth,height=0.2\textheight]{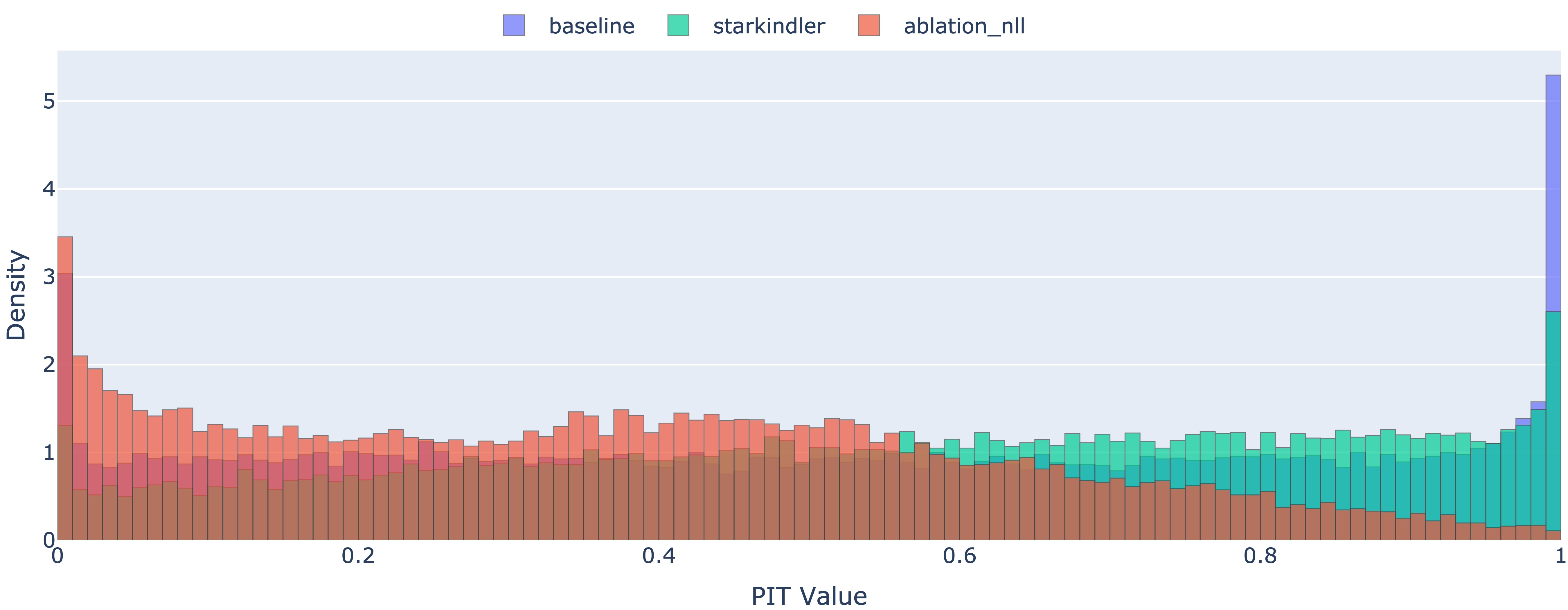} 
        \caption{PIT Histogram}
        \label{fig:fig2c}
    \end{subfigure}
    \vspace{-10pt} 
    \caption{Comparison of Baseline, Starkindler and the Ablation models}
    \label{fig:all}
\end{figure}

\vspace{-5pt} 
\textbf{Improved Prediction Accuracy}: Starkindler gets lower redshift prediction MSE and MAE errors as well as tighter error bounds as compared to the SDSS baseline (Table\ref{tab:performance}), while significantly outperforming the ablation model. Starkindler gets comparable contour plots as compared to the SDSS baseline, while the ablation model shows clear bias and non uniformity in its contour plots. Starkindler also reduces the outlier prediction in higher $z$ values in general, while particularly reducing the outliers due to under prediction of $z$ (Figure \ref{fig:fig2a}).

\textbf{Improved Uncertainty Calibration}: As compared to the baseline, Starkindler exhibits superior ECE values, better aligned calibration plots where alignment with the diagonal indicates better coverage (Figure \ref{fig:fig2b}),  and a more uniform PIT histogram (Figure \ref{fig:fig2c}), indicating improved calibration over ablation\_nll model. Interestingly, the ablation\_nll model gets a lower ECE score but a more skewed PIT plot, indicating a degree of overestimation of uncertainties. No model gets absolutely uniform PIT histograms, indicating scope for calibration and future improvement. Starkindler shows better coverage than the baseline and the ablation model throughout.

\section{Discussion}
\vspace{-5pt} 
Starkindler highlights the benefit of using observed error in the redshift data,  $zErr_{spec}$, in the training objective of ML models.

\textbf{General Improvements to Prediction Accuracy and Calibration}: Starkindler shows better redshift prediction accuracy with tighter error bounds and a lower outlier prediction rate as compared to the baseline. The ablation model which has the same model and training hparams and just differs in the training objective, gets poorer accuracy, worse outliers and poor calibration. While worse in predictive accuracy, the ablation model does get the lowest ECE scores which indicates the propensity of generic heteroskedastic models to overestimate uncertainty. This highlights the regularising effect played by inclusion of aleatoric uncertainty as provided by the Starkindler objective.

\textbf{Scientific Implications}: In contrast with existing probabilistic models for Redshift PDF prediction, Starkindler directly predicts the mean and standard deviation of the redshift for each photometric observation, while accounting for aleatoric uncertainty in the training data. This provides a more interpretable framework for modeling redshift errors which are critical for downstream tasks like weak lensing\cite{10.1093/mnrasl/slw201}\cite{10.1093/mnras/sty957}.

\section{Future Work}
\vspace{-10pt} 
\textbf{Scope for improvement of existing models}: Starkindler's training objective is model agnostic and can be integrated with existing well performing models, with minimal changes to their design.
\textbf{Hyper Parameter Tuning}: Hyper parameter tuning may further improve model performance.
\textbf{Beyond Redshift Estimation}: Starkindler can also be extended beyond redshift estimation to other astrophysical measurements where inherent uncertainties estimates are available.

\end{document}